\documentclass{amsproc}

\usepackage{latexsym}

\newcommand{\szego}{Szeg\"o }

\newcommand{\inv}{^{-1}}
\newcommand{\kahler}{K\"ahler }
\newcommand{\sqrtn}{\sqrt{N}}
\newcommand{\wt}{\widetilde}

\newcommand{\PP}{{\mathbb P}}
\newcommand{\R}{{\mathbb R}}
\newcommand{\C}{{\mathbb C}}

\newcommand{\Z}{{\mathbb Z}}

\newcommand{\CP}{\C\PP}

\newcommand{\D}{{\mathbf D}}
\renewcommand{\H}{{\mathbf H}}
\newcommand{\half}{{\frac{1}{2}}}

\renewcommand{\phi}{\varphi}

\newcommand{\ga}{\gamma}

\newcommand{\De}{\Delta}

\newtheorem{theo}{{\sc Theorem}}[section]

\title[From random polynomials to symplectic  ]
{From random polynomials to symplectic geometry }

\author{Steve Zelditch}
\address{Department of Mathematics, Johns Hopkins University, Baltimore,
MD
21218, USA}
\email{ zelditch@math.jhu.edu} 

\thanks{Research partially supported by NSF grant  \#DMS-9703775 .}

\date{June 7, 2000}

\begin{document}

\begin{abstract} We review some recent results on random polynomials and their generalizations
in complex and symplectic geometry.  The main theme is the universality of statistics of zeros and
critical points of (generalized) polynomials of degree $N$ on length scales of order $\frac{D}{\sqrt{N}}$
(complex case), resp. $\frac{D}{N}$ (real case). 
\end{abstract}

\maketitle

\section{Introduction} 

This is a short survey of  some  results of P. Bleher, J. Neuheisel,  B. Shiffman and the author on random polynomials and
their generalizations to holomorphic (and almost-holomorphic) sections of ample line bundles, mainly following   \cite{BSZ1, BSZ2, BSZ3,N,  SZ1, SZ2, SZ3, SZ4, Ze1, Ze2}.   Motivation to study random polynomials and their generalizations in geometry comes from several sources:
\medskip

\begin{itemize} 

\item Classical Analysis: Value distribution theory of polynomials and analytic functions is a classical topic.
Computable examples may exhibit non-generic patterns of zeros (or other values)  and one would like to understand the typical distribution.  One forms ensembles of analytic functions by defining  the coefficients to be independent random variables with a given distribution. One can then study  expected behaviour, almost sure behaviour and so on (see e.g. \cite{K, LO, O}).  

\item PDE: Spherical harmonics of degree $N$ are examples of eigenfunctions of the Laplacian on a compact Riemannian manifold. They
are restrictions to the sphere $S^m$ of homogeneous  harmonic polynomials of degree $N$ on $\R^{m + 1}$. 
One
would like to know about  nodal lines, critical points, sup norms (etc.) of general Laplace eigenfunctions.  Studying features of random spherical harmonics gives insight into 
the `typical' properties of eigenfunctions and avoids pathologies such as occur in  \cite{L, JN}. Analogues 
of spherical harmonics of degree $N$ on general compact Riemannian manifolds are linear combinations of eigenfunctions
$\Delta \phi_{\lambda}= \lambda^2 \phi_{\lambda} $ with $\lambda \in [N \log N, (N + 1) \log (N + 1)].$ 
 References on random spherical harmonics include  \cite{Be, N, V, Ze1}); for random combinations
of eigenfunctions, see \cite{Ze2}.

\item Quantum Chaos: Eigenfunctions of quantum chaotic Hamiltonians are well modeled by 
random polynomials in regard to
distribution of zeros or  critical points, to sizes (sup-norms or $L^p$-norms), to quantum expectation values, and in
other respects (see, among others, \cite{ABST, BBL, HKZ, NV, Ze1, Ze2}).  This is analogous to the similarity between eigenvalues of random matrices and eigenvalues of quantum chaotic systems.

\item Algebraic Geometry: Holomorphic sections $s \in H^0(M, L^N)$ of the $N$th power of an ample line bundle 
$L \to M$ over
a \kahler manifold $(M, \omega)$ are quite analogous to homogeneous polynomials of degree $N$, and coincide with 
such polynomials when $M = \CP^m, L = {\mathcal O}(1)$ ( $\CP^m = $ complex projective $m$-space,
${\mathcal O}(1)$ is the hyperplane bundle (cf. \cite{GH})). The simultaneous zero set $Z_{s_1, \dots, s_k}$ of $k$ holomorphic sections defines a codimension $k$  algeraic submanifold of $M$;  one would like
to know the `almost sure' properties of such a submanifold.

\item Symplectic Geometry: Almost-holomorphic sections $s \in H^0_J(M, L^N)$ of ample line bundles $L \to M$
over almost-complex symplectic submanifolds $(M, J, \omega)$  in the sense of \cite{DON, BG} are very similar to holomorphic sections in the complex case.
Under (difficult) transversality conditions, they have  applications in symplectic geometry analogous to those
in algebraic geometry.

\end{itemize}
\medskip

In this article, we will refer to the setting of holomorphic (or almost-holomorphic) sections as the `complex case', and the
setting of eigenfunctions of Laplacians as the `real case'. The complex wave functions live on  phase space
while the real eigenfunctions live on configuration space. 
The main theme of our work has been the universality of statistics of zeros and critical points of polynomials
of large degree $N$  on small length
scales (of order $\frac{D}{\sqrt{N}}$ in the complex case, resp. $\frac{D}{N}$ in the real case). We have only considered compact manifolds, and only
Gaussian or spherical measures on their spaces of polynomials.
Two  universality classes have emerged: (i) the
`Heisenberg class' in the complex  case (connected to the Heisenberg group), and (ii) the `Euclidean class' in real case
(connected to the Euclidean motion group). It should be mentioned that the `real case' has many other meanings in
the literature on random polynomials (e.g. real polynomials and their real zeros).

\section{Mathematical Tools}

We give a quick summary of some basic tools and methods that are used below. We will concentrate on general ideas
and refer to \cite{SZ1, SZ2, BSZ3} for detailed expositions. 

\subsection{Vector spaces of large dimension}

In defining our ensembles of polynomials, we will be dealing  with a sequence $( {\mathcal H}_N, \langle, \rangle_N)$ of Hilbert spaces of increasing dimension  $d_N = dim {\mathcal H}_N$, where $N$ is the `degree of the polynomial', 
 a large integral (semiclassical) parameter.  The dimension is given by a (Hilbert) polynomial
of the form  $d_N \sim a_0 N^m$ in the complex case (with
$m = dim_{\C} M$) and by a function of polynomial growth $d_N \sim  a_0 N^{m - 1}$ in the real case (with $m = \mbox{dim}_{\R} M$). 

In our applications, the spaces ${\mathcal H}_N$ will be one of the following.

\subsubsection{ Spherical harmonics and real eigenfunctions}

 We denote by $\Delta$ the standard Laplacian on $S^m$ and by ${\mathcal H}_N (S^m)$ the
space of spherical harmonics of degree $N$ on $S^m.$  They are the eigenfunctions of $\Delta$ of eigenvalue
$\lambda_{m,N} = N ( N + m - 1)$ and form a real vector space of dimension $d_{m, N} = \frac{2N + m -1}{N + m -1}
\left( \begin{array}{l} N + m -1 \\ \\ m -1 \end{array} \right)$. 

More generally, we may consider the Laplacian $\Delta$ of any compact Riemannian manifold $(M, g)$. In place of
 spherical harmonics of degree $N$, we partition the spectrum of $\sqrt{\Delta}$ into intervals $[N  \log N, (N+ 1) \log (N + 1)]$ (the reason for the longer length is given in \cite{Ze2}), and let ${\mathcal H}_N$ denote the span of the eigenfunctions with eigenvalues in the $N$th interval.
Linear combinations of such eigenfunctions are of course  not eigenfunctions of $\Delta$, but they behave
like polynomials of degree $N$ and  their random linear combinations give a  replacement for
random spherical harmonics.  Their use for modelling quantum ergodic and quantum mixing eigenfunctions is discussed \cite{Ze2, HKZ}. 

\subsubsection{Holomorphic sections of positive line bundles}

For any \kahler manifold $(M,\omega)$ of complex dimension $m$ there exists  a holomorphic hermitian line bundle $(L, h) \to (M, \omega)$ whose Ricci curvature $Ric(h) = \omega$. $L$ is called {\it positive} since it posseses a metric of positive curvature, i.e.  $\omega(X, JY)$ defines a Riemannian metric.  We denote by $L^N$ the $N$th power of $L$ and by  $H^0(M, L^N)$ the space of 
holomorphic sections. Its dimension $d_N = dim H^0(M, L^N)$ is given by the Hilbert polynomial $d_N = \frac{c_1(L)^m}{m!}
N^m + \cdots$ for sufficiently large $N$, where $\cdots$ represent the lower order terms. We equip $M$ with the volume 
form $dV_{\omega} = \frac{\omega^m}{m!}$ and $H^0(M, L^N)$ with the inner product $||s||^2 = \int_M h(s(z), s(z)) dV.$ 
For background we refer to \cite{GH}.

In the simplest case of Riemann surfaces,  
examples include:
\medskip

\begin{itemize}

\item $M = \CP^1, L = {\mathcal O}(1), L^2 = T \CP^1, h = h_{FS}, \omega = \omega_{FS}$ (Fubini study hermitian
metric and curvature (1,1)-form).  That  $(\CP^1, \omega_{FS})$ is positively curved in the usual Riemannian sense
is equivalent to positivity of $T\CP^1$.  $H^0(\CP^1, {\mathcal O}(2 N))$ may be interpreted as the 
space of holomorphic vector fields of type $(\frac{\partial}{\partial z})^N$. More simply put,
sections are homogeneous holomorphic polynomials $s = \sum_{j = 0}^N c_k z_0^k z_1^{N - k}$ of degree $N$ in two complex variables. Such polynomials are known as the $SU(2)$-ensemble.

\item $M = {\bf H}^2/\Gamma, L = T^*M, h_{FS} = $ hyperbolic metric. Hyperbolic surfaces are negatively curved in
the Riemannian sense, so their tangent bundles are negatively curved, and  their co-tangent bundles are positively curved.
Holomorphic sections are holomorphic differentials of type $(dz)^N$.  

\item $M = \C/ \Z^2, \omega_0 = dz \wedge d\bar{z}$. The complex torus is flat in the Riemannian sense, so neither its tangent nor cotangent bundles
are positively curved. The `quantizing line bundle' with curvature $\omega_0$ is rather the bundle $\Theta$ whose
sections are the classical theta-functions. The sections of $\Theta^N$ are known as theta-functions of level $N$.

\end{itemize}

\subsubsection{Almost holomorphic sections}

Symplectic almost-complex manifolds $(M, J, \omega)$ possess a similar but analytically more
complicated geometric quantization as spaces $H^0_J(M, L^N)$ of `almost-holomorphic' sections. They are
defined by a $\bar{D}$ complex over the $S^1$-bundle $X$ due to Boutet de Monvel -Guillemin \cite{BG}.  What
we need to know about these spaces is that their orthogonal projectors $\Pi_N$ have the same scaling asymptotics
as in the complex case, if one works in suitable (Heisenberg) coordinates \cite{SZ2}.

\subsection{Gaussian measures and spherical measures}

We will restrict attention to two related ensembles:
\begin{itemize}

\item  {\it Gaussian ensembles}: We fix an orthonormal basis $\{f_j\}$ of ${\mathcal H}_N$ and write
functions as orthonormal sums $f = \sum_{j = 1}^{d_N} c_j f_j$. We then define the (complex) Gaussian measure by
$$\gamma = e^{- ||f||^2} {\mathcal D}f,\;\; \mbox{i.e.} \;\; \gamma = e^{- |c|^2} dc.$$
More generally we fix a symmetric matrix $\Delta$ on $\C^{d_N}$  with positive (semi-)definite imaginary part
and define:
\begin{equation}\ga_\De = \frac{e^{- \langle\De\inv
c,c\rangle}}{(2\pi)^{p/2}\det\De}
dc\,,\label{gaussian}\end{equation}
Gaussian ensembles come in both real and complex flavors. In the real case, the exponents acquire factors of $1/2$
and the denominator acquires a square root.

\item {\it Spherical ensembles}: We denote by $S{\mathcal H}_N = \{f \in {\mathcal H}_N: ||f|| = 1\}.$ We then
equip $S{\mathcal H}_N$ with the uniform (Haar) probability measure $\nu_N.$

\end{itemize}

We will denote the expected value of a random variable $X$ with respect to the Gaussian ensemble by ${\bf E}_{\gamma}(X)$
(resp. ${\bf E}_{\nu}(X)$ for the spherical ensemble).

These two ensembles are {\it equivalent} in the sense that the two
large dimension limits give equivalent results when scaled properly. A more precise formulation goes as follows:
\medskip

{\it \label{spherical-vs-gaussian}
 Let $T_N: \R^{d_N} \to R^k$, $N=1,2,\dots$, be a sequence of linear maps,
where $d_N
\to \infty$.  Suppose that $\frac{1}{d_N}T_N T_N^*
\to
\Delta$. Then
$T_{N*} \nu_{d_N} \to \gamma_{\Delta}$. }
\medskip

\subsection{Sequences of random polynomials}

We are often interested in  sequences of polynomials $\{s_N\}$ chosen independently and at random
from ${\mathcal H}_N$ from either a Gaussian or spherical ensemble. We thereform form
the product probability space $({\mathcal H}_{\infty}, \mu_{\infty})$, defined by
$${\mathcal H}_{\infty} = {\mathcal H}_1 \times {\mathcal H}_2 \times \cdots \times {\mathcal H}_N \times \cdots,\;\;\; \mu_{\infty} = \times_{N = 1}^{\infty} \mu_N, \;\; (\mu_N = \gamma_N\;
\mbox{or}\; \nu_N). $$
When we say that a sequence of polynomials $\{s_N\}$ of increasing degerees does something almost surely, we mean
that the set of such sequences has measure one in this product ensemble.

\subsection{\szego kernels}

Our results depend on the fact that certain statistical properties of polynomials can be expressed in terms of 
the reproducing kernels $\Pi_N(x,y)$ (orthogonal projections) of the Hilbert spaces ${\mathcal H}_N$.
They are known as Szego kernels, and  are essentially the same as the  
`coherent states' of the physics literature. The local structure of the \szego kernel is given by the following
scaling asymptotics:

\begin{theo} As $N \to \infty$, we have:

\begin{itemize}

\item \cite{BSZ2, SZ2} Complex case: 
$$  \Pi_N(z_0 + \frac{u}{\sqrt{N}}, \frac{\theta}{N}, z_0 + 
\frac{v}{\sqrt{N}}, \frac{\phi}{N})
\sim \frac{1}{\pi^m} e^{i (\theta - \phi)}
e^{u \cdot \bar{v} -\half (|u|^2 + |v|^2)} \{1 + \frac{1}{\sqrt{N}}
p_{1}(u,v; z_0)
+ \cdots \}\,.$$
The leading order term is the \szego kernel for the reduced Heisenberg group, whence the name `Heisenberg class.'

\item Real case: $$ N^{-m + 1} \Pi_N(x_0 + \frac{u}{N},  x_0 + 
\frac{v}{N})
\sim \Gamma(\frac{m - 1}{2}) (\frac{|u - v|}{2})^{\frac{m - 2}{2}} J_{\frac{m - 2}{2}}(|u - v|) + \cdots \,,$$
where $J_{\nu}$ is the Bessel function of the first kind of order $\nu$ (whence the name `Euclidean class'.) 
\end{itemize}
\end{theo}

The proof of the scaling asymptotics in the complex holomorphic case \cite{BSZ2} is based on the Boutet-de-Monvel- Sjostrand
parametrix for the \szego kernel, which is valid for positive line bundles. A similar parametrix was constructed
in the symplectic almost-complex case \cite{SZ2}, and the  scaling asymptotics were derived from it. In the real case
of $S^m$,
the scaling asymptotics are closely related to the `Mehler-Heine formula'.  The
terms `Heisenberg class' and `Euclidean class' suggest infinite dimensional Gaussian ensembles related to 
representations of the Heisenberg and Euclidean motion groups.

\section{Distribution of zeros and critical points}

We now state some results on the zeros and critical points of random generalized polynomials.   Let  $(s_1, \dots, s_k) \in H^0(M, L^N)^k$ or let $(s_1, \dots, s_k) \in {\mathcal H}_N^k$
in the real case.  
When $k = 1$ we omit the subscript. We will use
the following notation.
\begin{itemize}

\item $Z_s = \{x : s(x) = 0\}$ denotes  the zero set of $s$, and 
$|Z_s|$ denotes the $m - k$-submanifold Riemannian volume density induced by $\omega$ in the complex case or by
$g$ in the real case. We further denote by $||Z_s||$ the mass of $|Z_s|$ and define the probability measures  $\tilde{Z}_s = \frac{|Z_s|}{||Z_s||}.$ When one takes $m$ sections (or functions) in dimension
$m$, then the simultaneous zeros are almost surely a discrete set and the measure $\tilde{Z}_s$ is the normalized
sum of delta-functions at the zeros.

\item $C_s =\{z: \nabla s(z) = 0\}$ denotes the critical point set of $s$. In the holomorphic case, 
 $\nabla$ is the holomorphic connection compatible with $h$.   We note that $C_s$ is almost surely a discrete set. We define 
$|C_f| = \sum_{z_j: \nabla s(z_j) = 0} \delta (z_j),$ $ ||C_s|| = \# C_s,$ and $\tilde{C}_s = \frac{|C_s|}{\# C_s}.$

\item By the density of zeros at degree $N$  we mean the coefficient $K^N_{1,k}(z)$ of the  measure $K^N_{1,k}(z) dV = {\bf E} |Z_s|,$ i.e. 
$\int_M \phi {\bf E} |Z_s| = {\bf E} \int_M \phi |Z_s|$ for $\phi \in C(M).$ Similarly, we denote by $K^{crit, N}_{1}(z)$
the density (relative to the given volume form) of ${\bf E}\tilde{C}_s.$

\item More generally, we define the pair correlation densities of zeros (resp. critical points) by
$K^N_{2,k}(z^1, z^2) dV = {\bf E} (|Z_s| \times |Z_s|),$ resp. $K^{crit, N}_{2,k}(z^1, z^2) dV = {\bf E} (|C_s|\times |C_s|).$ They are densities of measures on $M \times M.$ Roughly speaking, the two-point correlation gives the
probability density of finding a pair of zeros (or critical points) at $(z^1, z^2).$ More generally, there are 
n-point correlation functions, but for simplicity we only consider $n = 1, 2.$

\end{itemize}

\subsection{Statement of results} 

We have results on several levels: expected values, almost sure behaviour, and scaling asymptotics. The following
theorems are valid on any complex or almost-complex symplectic manifold $(M, \omega)$, equipped with a hermitian complex line
bundle of curvature $\omega.$ 

\begin{theo}   In the complex case, the  density of zeros, resp. critical points, satisfies:

\begin{itemize} 

\item \cite{SZ1} Complex zeros: $K^N_{1,k}(z) dV = \omega^k + O(\frac{1}{N}).$

\item \cite{SZ3} Complex critical points: There exists a universal constant $\gamma_m$ depending only on the dimension such that 
$ K^{crit, N}_1(z) dV = \gamma_m \frac{\omega^m}{m!} N^m + O(N^{m - 1}),$   In particular, the expected number
of critical points is given by ${\bf E} \# C_s =  \gamma_m Vol_{\omega}(M) N^m + O(N^{m - 1}).$ 
Here, $Vol_{\omega}(M) = \frac{c_1(L)^m}{m!}$ is the volume of $(M, \omega)$. The density
of critical points is therefore universal.

\end{itemize}

\end{theo}

Thus, zeros and critical points tend to concentrate in regions of high curvature. Similar results should hold in the real case. In the case
of $S^m$, such density results are obvious since they must be rotationally invariant. 

\subsection{Almost sure distribution of zeros}

A deeper question on distribution of zeros is whether sequences of  individual sections tend to have equidistributed zeros.

\begin{theo} We have:

\begin{itemize}

\item \cite{SZ1} For a random sequence in the complex codimension $k$ case, $\{s_N\}$, $\tilde{Z}_{s_N} \to \omega^k$ almost surely.

\item \cite{N} For a random sequence of spherical harmonics on $S^m, m \geq 6$, we have $\tilde{Z}_{s_N} \to dvol$ almost surely. The same is true in (Cesaro) mean for dimensions $< 6.$

\end{itemize}
\end{theo}

\subsection{Universality and scaling of correlations}

The next level of results concerns the statistics of zeros and critical points on the length scale $\frac{D}{\sqrt{N}}$
(complex case) or $\frac{D}{N}$ (real case).  For brevity we only describe the results in the complex case.
Upon magnifying a small ball $B_{\frac{D}{\sqrt{N}}}(z_0)$ around an arbitrary point $z_0$ of this radius, one 
 loses track of the specifics of the geometrical setting and obtains universal
limiting correlations. More precisely, such a universal limit occurs if one chooses the coordinates properly.

\begin{theo}\label{usls} Scaling limits of correlations of zeros and critical points are universal.  That is:
$$\frac{1}{ N^{2k}} K_{2k}^N\left(\frac{z^1}{\sqrtn}, 
\frac{z^2}{\sqrtn}\right) \to K_{2km}^{\infty}(z^1,z^2)$$ 
in the sense of measures. 
\end{theo}

The  limit correlation function denoted $K_{2km}^{\infty}(z^1,z^2)$ is unique within a universality class for
each statistic under consideration. In the complex and almost-complex cases, there exists a limit correlation function
$K_{2km}^{\C, zeros, \infty}(z^1,z^2)$ of zeros, 
  and  another limit correlation function  $K_{2km}^{\C, crits, \infty}(z^1,z^2)$ for critical points.  
There are analogous results in the real case. The limit correlations are explicitly computable. In \cite{BSZ1}
and elsewhere we give explicit formulae for low
values of $m$ and graph the results. The result for zeros in dimension one agrees with the formula of Hanny \cite{Ha}
on $\CP^1$, as it must since the result is universal. We now briefly describe the elements of the proof. The details differ, but the principles are the same,
for  complex and real cases, and for zeros or critical points. Hence we concentrate on zeros in the complex case.

\subsubsection{Step One: Relating correlations and joint probability distributions}

Following an idea due originally to Kac  and Rice  in the case of real polynomials of one variable, we
express the correlation measures in terms of the joint probability 
distribution  of the random variables 
$s(z^1),\dots,s(z^n),
\nabla s(z^1),\dots,\nabla s(z^n)$.
This JPD is defined by $$\wt\D_{z}^N:=
\wt D_{n}^N(x,\xi,z)dxd\xi=(J_z)_*\nu_N\,,
\;\;J_{z}(s) = (s(z), \nabla s(z)),$$ i.e. it is the push-forward of the Gaussian measure $\gamma_N$ under the  linear jet  map $J(z)$.  

The desired expression for correlations in terms of the JPD is given by the following generalization of the Kac-Rice formula \cite{ K} to the geometric setting of this article: 
\begin{theo}\label{introdn} \cite{BSZ1, BSZ3, SZ2} In the case of correlations of complex zeros, we have:
$$K_{2k}^N(z)=
\int d\xi\,\wt D_n^N(0,\xi,z)\prod_{p=1}^n
\det(\xi^{p}\xi^{p*})\,.$$
\end{theo}

Analogous formulae exist for critical points, but involve the jet maps $(\nabla s(z), \nabla^2 s(z))$. The
real case is similar to the complex case, but is  somewhat more complicated because $\Pi_N$ is oscillatory 
rather than exponentially decaying \cite{N}.  

\subsubsection{Step two: scaling asymptotics of the JPD}

The JPD  is  a (generalized) 
Gaussian measure  on the complex vector space of 1-jets:
$\wt\D_{z}^N=\ga_{\De^N(z)}\,,$ 
where  the covariance
matrix $\De^N(z) = \left(
\begin{array}{cc}
A & B \\
B^* & C
\end{array}\right)$ is given in terms of  the Szeg\"o kernel and its covariant
derivatives, as follows:
$$ \begin{array}{ll} A = \big( A^{p}_{p'}\big) =\frac{1}{d_N}\Pi_N(z^p,0;z^{p'},0)\,, & 
B = \big(B^{p}_{p'q'}\big)= \frac{1}{d_N}
\overline{\nabla}^2_{q'}\Pi_N(z^p,0;z^{p'},0)\,, \\ & \\
C = \big(C^{pq}_{p'q'}\big) = 
 \frac{1}{d_N}
\nabla^1_q\overline{\nabla}^2_{q'}\Pi_N(z^p,0;z^{p'},0)\,, &  p,p'=1,\dots,n, \quad q,q'=1,\dots,2m\,.\nonumber
\end{array}$$
Here, $\nabla^1_q$, respectively $\nabla^2_q$, denotes
the differential operator on $X\times X$ given by applying
$\nabla_q$ to the first, respectively second, factor.  The link between the JPD and the Szego kernel stems ultimately from the fact that $\Pi_N$ is the covariance matrix of $\gamma$ on
${\mathcal H}_N.$  Using the scaling asymptotics of the \szego kernel we obtain (in the complex case):

\begin{theo}\label{usljpd} {\rm (\cite{SZ2}, Theorem 5.4)} With the same notations and assumptions, we have:
$$\wt \D^N_{(z^1/\sqrtn,\dots, z^n/\sqrtn)} \longrightarrow \D^\infty
_{(z^1,\dots,n^n)}= \ga_{\De^\infty(z)}$$   where
$\D^\infty_{(z^1,\dots,z^n)}$ is a universal Gaussian measure, and $\De^N(z/\sqrtn)\to \De^\infty(z)$.
\end{theo}

The covariance matrix $\De^\infty$ is given in terms of the \szego kernel for the
Heisenberg group:
\begin{equation}\label{delta}
\De^\infty(z)= \frac{m!}{c_1(L)^m}\left(
\begin{array}{cc}
A^\infty(z) & B^\infty(z) \\
B^{\infty}(z)^* & C^\infty(z)
\end{array}\right)\,,\end{equation} where
\begin{eqnarray*} A^\infty(z)^p_{p'} &=& \Pi_1^\H(z^p,0;z^{p'},0)\,,\\
B^\infty(z)^{p}_{p'q'} &=&\left\{\begin{array}{ll}
(z^p_{q'}-z^{p'}_{q'}) \Pi_1^\H(z^p,0;z^{p'},0) \  &\mbox{for}\quad  1\le
q\le m\\ 0 & \mbox{for}\quad  m+1\le q\le 2m\end{array}\right.\ ,\\
C^\infty(z)^{pq}_{p'q'} &=&\left\{\begin{array}{ll}
(\delta_{qq'} + (\bar z^{p'}_q
-\bar z^p_q) (z^p_{q'}-z^{p'}_{q'}))\Pi_1^\H(z^p,0;z^{p'},0) \ 
&\mbox{for}\quad  1\le q,q'\le m\\ 0 & \mbox{for}\quad  \max(q,q')\ge
m+1\end{array}\right.\ .
\end{eqnarray*}

The scaling limit thus gives the correlations between zeros in the `Heisenberg ensemble', an infinite dimensional
Gaussian ensemble. 
In the real case (i.e. $S^m$), the limit correlations coincide with those in an ensemble related to
the Euclidean motion group, and involving the Bessel kernel \cite{N}. 

\subsection{Hole probabilities}

Another application of the scaling applications is to `hole probabilities', i.e. the 
probability that a ball $B_{r}(z_0)$ of radius
$r$ around a point $z_0 \in M$ is zero-free.  The following result combines   our scaling asymptotics
and Sodin's reformulation (and substantial simplification) of Offord's large deviations results on hole probabilities
for entire analytic functions in the plane  \cite{O, So}. It is based on the Poincare-Lelong formula, so at this time of
writing it has
only been proved for one holomorphic section, and  no comparable results have been proved for random spherical harmonics.

\begin{theo} Let $P_N(D; z_0) = Prob\{s \in H^0(M, L^N): Z_s \cap B_{\frac{D}{\sqrt{N}}}(z_0) = \emptyset\}.$ Then there
exists positive constants $C_1, C_2$ such that, for
any $N$,  $P_N(D; z_0) \leq C_1 e^{-C_2 D^2}.$\end{theo}

\section{Quantum ergodicity and random waves} 

We end with a brief discussion of the connection between random polynomials and quantum chaos. 
The intutitive idea that quantum chaotic eigenfunctions
should resemble `Gaussian random waves' seems to have been first  suggested by M. V. Berry \cite{B}. 
A precise formulation of
this random wave model and some numerical  results  are given in  \cite{ABST}  \cite{HR}.

Random spherical harmonics on $S^m$, or random combinations of eigenfunctions of Laplacians on general compact
Riemannian manifolds as described above, provide a rather different random wave model for quantum chaotic eigenfunctions.
To motivate the model, we recall that the diagonal sums of squares
\begin{equation}\label{QE} S_p(\lambda) = \sum_{j: \lambda_j \leq \lambda} |(A \phi_j, \phi_j) - \bar{\sigma_A}|^p,\;\;\;
(\mbox{with}\;\;\bar{\sigma_A} = \int_{S^*M} \sigma_A d \mu), \end{equation}
and their  off-diagonal analogues, are used to characterize eigenfunctions as quantum ergodic, quantum mixing and so on.  
Here, $A$ is an observable (zeroth order pseudodifferential operator), $\sigma_A$ is its principal symbol. 
For quantizations of ergodic systems, $S_p(\lambda) \to 0$ as $\lambda \to \infty$,
 and it is of some interest
to measure the rate and to relate it to the classical dynamics. In the  random spherical harmonics model one has  the 
following rate (for similar results see \cite{Ze2, SZ1}):

\begin{theo} (\cite{Ze1}, Lemma (2.15))  Let $\{\phi_{N j}\}$ be a random orthonormal basis of spherical harmonics of $S^2$. Then:
$E (S_2(N)) = \frac{1}{N}[\frac{1}{vol(S^*S^2)} \int_{S^*S^2}|\sigma^{ave}_A(\zeta) - \bar{\sigma_A}|^2 d\mu(\zeta)] + O(1/N^2). $ \end{theo}

The optimist may conjecture a similar rate for quantum chaotic eigenfunctions. In \cite{Ze1} it is further proved
that almost all orthonormal bases  of spherical harmonics are quantum ergodic, and this was improved to quantum unique ergodicity by VanderKam \cite{V}.  On the level of quantum mixing, however, random spherical harmonics do not provide
a good model, but random combinations of Laplace eigefunctions on generic Riemannian manifolds do; that was the
motivation for studying the model in \cite{Ze2}.   Quantum mixing
systems involve off-diagonal sums like (\ref{QE}) with constant gaps between eigenvalues. 
Conversely, if the eigenfunctions satisfy  (\ref{QE}) and the analogous off-diagonal estimates for all gaps, then
the system is classically mixing.   Hence the random wave model is not a good model for ergodic systems which fail
to be mixing (cf. \cite{HKZ}).

A further relation between Gaussian random waves and quantum chaotic eigenfunctions does not seem to have been explored,
even numerically. Let $\{\phi_{\lambda}\}$ be eigenfunctions of a quantum chaotic system, and consider the local
rescaling $\phi_{\lambda}(x_0 + \frac{u}{\lambda}).$ The rescaled eigenfunction is an eigenfunction of the rescaled
operator, which is asymptotically Euclidean. Hence  $\phi_{\lambda}(x_0 + \frac{u}{\lambda})$ can be expanded asymptotically in terms of plane waves, and one might ask how the frequencies are distributed.

\end{document}